\documentclass[useAMS]{mn2e}

\usepackage{amssymb,amsmath,psfig,times,url}
\def\gsim{ \lower .75ex \hbox{$\sim$} \llap{\raise .27ex \hbox{$>$}} }
\def\lsim{ \lower .75ex\hbox{$\sim$} \llap{\raise .27ex \hbox{$<$}} }

\def\ergs{{\rm\thinspace erg \thinspace s^{-1}}}

\def\sc{Schwarzschild}
\def\beq{\begin{equation}}
\def\eeq{\end{equation}}

\def\sc{Schwarzschild}

\title[Blazar  at $z>4$]
{Blazar candidates beyond redshift 4 observed by {\it Swift}} 
\author[T. Sbarrato et al.]
{T. Sbarrato$^{1,2}$\thanks{Email: tullia.sbarrato@brera.inaf.it}, 
G. Ghisellini$^2$, 
G. Tagliaferri$^2$,
L. Foschini$^2$,
M. Nardini$^3$, F. Tavecchio$^2$,   
\newauthor{N. Gehrels$^4$}
\\ \\
$^1$Dipartimento di Scienza e Alta Tecnologia, Universit\`a dell'Insubria, 
Via Valleggio 11, I--22100 Como, Italy\\
$^2$INAF -- Osservatorio Astronomico di Brera, Via Bianchi 46, I--23807 Merate, Italy\\
 $^3$Dipartimento di Fisica G. Occhialini, Universit\`a di Milano Bicocca, 
 Piazza della Scienza 3, I--20126 Milano, Italy\\
$^4$ NASA--Goddard Space Flight Center, Greenbelt, Maryland 2077, USA
}

\begin{document}  

\maketitle

\begin{abstract}
We have selected SDSS J222032.50+002537.5 and SDSS J142048.01+120545.9
as best blazar candidates out of a complete sample 
of extremely radio--loud 
quasars at $z>4$, with highly massive black holes. 
We observed them and a third serendipitous candidate with similar features (PMN J2134--0419) 
in the X--rays with the {\it Swift}/XRT telescope, to confirm their blazar nature. 
We observed strong and hard X--ray fluxes 
(i.e.\ $\alpha_X\lsim0.6$, where $F(\nu)\propto\nu^{-\alpha_X}$ in the 0.3--10~keV observed energy range, 
$\sim$1--40~keV rest frame) 
in all three cases. 
This allowed us to classify our candidates as real blazars, being characterized 
by large Lorentz factors ($\sim13$) and very small viewing angles ($\sim3^\circ$). 
All three sources have black hole masses exceeding $10^9M_\odot$ and their 
classification provides intriguing constraints on supermassive black hole 
formation and evolution models. 
We confirm our earlier suggestion that there are different formation epochs 
of extremely massive black holes hosted in jetted ($z\sim4$) and non--jetted systems ($z\sim2.5$). 
\end{abstract}


\section{Introduction}
\label{intro}

Blazars are radio--loud Active Galactic Nuclei (AGN) with relativistic jets 
oriented along our line--of-sight (Urry \& Padovani 1995). 
The viewing angle $\theta_v$ is smaller or comparable to the jet beaming 
angle, hence the jet emission is strongly boosted because of relativistic effects. 
The strong boosting makes blazars particularly luminous and hence visible 
up to very high--redshift.
Nevertheless, before our works (Sbarrato et al.\ 2012; 2013a, b; Ghisellini et al.\ 2014), 
the only known blazars at $z>4$ were discovered serendipitously. 

In general, the Spectral Energy Distribution (SED) of a blazar is dominated 
by the jet emission over the whole energy range (from radio to $\gamma$--ray emission) 
and consists of two broad humps, produced by synchrotron (at low frequencies) 
and Inverse Compton (IC) emissions (at high frequencies).  
The SED peaks  shift at lower frequencies as the power of the 
jet increases, at least in the case of known blazars (Fossati et al.\ 1998; 
Donato et al.\ 2001; for a different interpretation see Giommi et al.\ 2012). 
Since at high redshift we expect to see only the most powerful objects, 
both this effect and the redshift itself 
will contribute to shift the observed SED 
at lower frequencies. 
This affects the 
search of high--$z$ candidates and their classification as blazars. 
In fact, at low redshift, complete surveys of blazars can be obtained with $\gamma$--ray 
instruments, like the Large Area Telescope (LAT) onboard the {\it Fermi Gamma--Ray 
Space Telescope} ({\it Fermi}; Atwood et al.\ 2009). 
The $\gamma$--ray emission is therefore the most common blazar fingerprint. 
At high redshift only the most powerful blazars are observable 
(see high--redshift tail of blazar samples, e.g.\ Ajello et al.\ 2009; Ackermann et al.\ 2011), 
i.e.\ the ones peaking 
below 100~MeV in the rest frame, and a factor $(1+z)$ less in the observer frame. 
Therefore these sources are more and more difficult to observe in the 
$\gamma$--ray band as the redshift increases. 

On the other hand, the high--energy component of a powerful blazar at high redshift 
can be observed in the X--ray range, mapping the SED just below the IC peak. 
In fact, the Burst Alert Telescope (BAT) onboard the {\it Swift} satellite (Gehrels et al.\ 2004) 
which is less sensitive than {\it Fermi}/LAT (Ajello et al.\ 2009) and has detected a much 
lower number of blazars, detected blazars up to redshift definitely higher than 
{\it Fermi}/LAT itself. 
A jetted source can then be classified as a blazar if it shows 
a hard X--ray spectrum [$\alpha_X\lsim0.6$, where $F(\nu)\propto\nu^{-\alpha_X}$]
along with an intense X--ray flux compared to the optical one. 
Unfortunately, 
the identified sources in BAT do not include blazars at $z>4$, 
and there is no other hard X--ray survey that can detect systematically blazars. 
This is why blazars at $z>4$ were only discovered serendipitously. 

To apply a more systematic approach to the search of high--redshift blazars, 
we first considered an optical flux limited sample covering a selected area of the sky. 
This is the Sloan Digital Sky Survey (SDSS; York et al.\ 2000). 
From this, we selected a sample of good blazars candidates (Sbarrato et al.\ 2013a), 
from which we already successfully classified our best candidate B2~1023+25 
thanks to X--ray observations performed by {\it Swift}/XRT and {\it NuSTAR} (Sbarrato et al.\ 2012; 2013b). 
The second most distant object of our selection, SDSS~J1146+403 at $z=5$ 
was confirmed as a blazar  too, with {\it Swift}/XRT observations (Ghisellini et al.\ 2014). 
The most distant blazar known (Q0906+6930; Romani et al.\ 2004; Romani 2006) 
was instead discovered serendipitously, and it is located at $z=5.47$.

The definition of ``blazar" is somewhat subjective. 
Generally speaking, a blazar is a source whose jet is seen at a ``small" 
viewing angle. 
But there is no exact definition of what ``small" means. 
Here we require $\theta_{\rm v}<1/\Gamma$ to classify a source as 
a blazar, mainly because in this way we can easily calculate the number 
of the parent population: for each source observed at $\theta_{\rm v}<1/\Gamma$, 
there must exist  other $2\Gamma^2$ sources pointing in other directions. 

In this work, we will discuss the classification of other three objects out of our sample, 
made possible by {\it Swift}/XRT observations. 
\S \ref{sel} describes the candidates final selection out of our sample of radio--loud quasars; 
\S \ref{data} provides the details of the observations performed by {\it Swift}; 
in \S \ref{jet} we discuss orientations and other features derived by fitting the observed data; 
\S \ref{comp} compares our results with what observed in the most known high--redshift blazars, 
while \S \ref{fi_sec} put our findings in the wider picture of extremely massive black holes 
in the early stage of our Universe; 
in \S \ref{concl} we summarize our findings. 

We adopt a flat cosmology with $H_0=70{\rm km s^{-1} Mpc^{-1}}$ and $\Omega_{\rm M}=0.3$.

\section{Candidate selection}
\label{sel}

We selected our candidates from a sample of $z>4$, extremely radio--loud 
quasars included in Sbarrato et al.\ (2013a). 
The selection we applied in our previous work had the main goal of collecting 
the highest--redshift jetted quasars that most likely have their jets directed near 
our line of sight, out of the Quasar Catalog (Schneider et al.\ 2010) from the 
7th SDSS Data Release (SDSS DR7). 
We built our sample of good candidates at $z>4$ by selecting those quasars 
with a large radio--loudness, i.e.\ $R=F_{\rm 1.4GHz}/F_{\rm2500\AA}>100$ 
(rest frame frequencies). 
Since the emission from the jet is relativistically beamed along the emitting 
direction, the dominance of its radio flux over the optical flux 
(which is quasi--isotropic, being emitted by the 
accretion disc)
is enhanced if the jet is directed towards our line of sight. 
On the other hand, 
the radio--loudness does not allow us to infer the exact 
orientation of the jet with respect to our line--of--sight, therefore preventing us 
from precisely classifying our candidates. 
We need another feature, which is provided by the X--ray flux and slope.
If pointing at us, in fact, the high energy component should dominate
the bolometric non--thermal luminosity, implying a strong X--ray flux
even in the classical [0.3--10] keV range, characterized by a hard spectrum
indicating that the peak of the $\nu F_\nu$ flux is at $\sim$MeV energies.
To this aim,
we selected our three best candidates to be observed by {\it Swift}:
they are the sources with the largest radio--loudness and the most intense
radio flux.

We selected the two best candidates from our restricted sample of 
19 SDSS quasars, i.e.\ SDSS J222032.50+002537.5 and SDSS J142048.01+120545.9
($z=4.205, R=4521$ and $z=4.034, R=1904$, respectively). 
Along with these sources, we included in our observations one extremely radio--loud and 
radio--luminous quasar not included in the SDSS DR7 Quasar Catalog, but coming from a compilation 
of $z>4$ quasars made by Djorgovski\footnote{\url{http://www.astro.caltech.edu/~george/z4.qsos}}, 
i.e.\ PMN J2134--0419 ($z=4.346, R=15843$). 
Even if not selected from the blazar candidates in Sbarrato et al.\ (2013a), 
PMN J2134--0419 is included in the field of the SDSS+FIRST survey. 
In fact, this object was photometrically observed in the SDSS, but it was not included 
in the SDSS DR7 quasar catalog because it did not fulfill the requirements of the 
color--based high--redshift quasar selection. 
Nevertheless, it is included in the same field as  SDSS J222032.50+002537.5 
and SDSS J142048.01+120545.9. 
This is important to correctly estimate the number of misaligned quasars inferred 
thanks to our observations (see \S\ref{jet}). 


The already mentioned successful classifications of B2~1023+25 and 
SDSS~J1146+403 at $z>5$ (Sbarrato et al.\ 2012; 2013b; Ghisellini et al.\ 2014; see \S\ref{intro}) 
strengthen the validity of our selection method, and encourage us to continue the 
detailed classifications of our candidates.

\begin{table*}
 \centering
  \begin{tabular}{lccccccccccc}
\hline
\hline
Name & ObsID & Exp & $N_{\rm H}$ &$F_{\rm norm}$ & $\Gamma_{\rm X}$ & $F_{0.3-10\rm keV}^{\rm obs}$ & Cash/d.o.f. &$v$\\ 
{}   & {} & [ks] & [$10^{20}$] &[$10^{-4}$] & {} & [$10^{-13}$] & {} & [mag] \\
\hline
SDSS J142048.01+120545.9    &00032625001  & 21.8 &1.74  &$3.3^{+2.5}_{-1.5}$ & $1.6\pm 0.3$ & $1.7$ & 64.13/65 & $20.43\pm0.10$\\
{}		      &00032625002  &	&	&	&	&	&	& \\
\hline
PMN J2134--0419   &00032624001   & 25.1 &3.34   &$2.7^{+2.5}_{-1.3}$ &$1.6\pm 0.3$ & $1.4$ &52.36/55 & $21.11\pm0.21$\\
{}		      &00032624002  &	&	&	&	&	&	& \\
{}		      &00032624003  &	&	&	&	&	&	& \\
\hline
SDSS J222032.50+002537.5    &00032626001  & 30.7 &4.41  &$1.1^{+1.1}_{-0.6}$ & $1.4\pm 0.3$ & $1.0$ & 53.30/51 &$21.16\pm0.18$  \\
{}		      &00032626002  &	&	&	&	&	&	& \\
{}		      &00032626003  &	&	&	&	&	&	& \\
\hline
\hline
\end{tabular}
\caption{
Summary of XRT and UVOT observations.
The column ``Exp" indicates the
effective exposure in ks, while $N_{\rm H}$ is the Galactic absorption
column in units of [$10^{20}$ cm$^{-2}$] from Kalberla et al. (2005). 
$F_{\rm norm}$ is the normalization flux at 1~keV in units of [$10^{-4}{\rm ph\,cm^{-2}s^{-1}keV^{-1}}$], 
$\Gamma_{\rm X}$ is the photon index of the power law model [$F(E)\propto E^{-\Gamma}$], 
$F_{0.3-10\rm keV}^{\rm obs}$ is the observed flux
in units of [$10^{-13}$ erg cm$^{-2}$ s$^{-1}$]. 
The next column indicates the value of the likelihood (Cash 1979) along with 
the degrees of freedom.
The last column reports the observed $v$ magnitude (not
corrected for absorption).
}
\label{xrt}
\end{table*}

\subsection{Black hole mass estimate}

The knowledge of the black hole mass is crucial to  study high redshift  blazars.
Because of the extremely high redshift, commonly used virial methods 
can rely only on CIV emission line. 
SDSS J222032.50+002537.5 and SDSS J142048.01+120545.9 have virial 
black hole mass estimates derived from CIV line luminosity, since they 
have been analyzed by Shen et al.\ (2011) as part of the SDSS DR7 quasar catalog. 
In the original sample of blazar candidates (Sbarrato et al.\ 2013a), the 
most distant quasars do not show any of the emission lines calibrated 
for the virial methods, since at $z>4.7$ even CIV falls outside the SDSS 
spectra energy range. 
Moreover, the SDSS does not provide a spectrum for PMN~J2134--0419. 

To derive consistently the $M_{\rm BH}$ for the whole Sbarrato et al. (2013a) 
sample, we applied a different method, that can be extended also to this source. 
We took advantage 
of the large power of our sources, implying a low
frequency peak of the synchrotron component, 
leaving the accretion disc flux ``naked" and thus dominating
the UV--optical--IR  emission.
With a good optical--IR data coverage, it is possible to directly fit the 
accretion disc emission with a simple model. 
Since in this kind of objects the CIV and/or the Ly$\alpha$ lines are clearly 
visible, we can expect that their accretion structures are radiatively efficient.  
We thus apply the simplest model for radiatively efficient disc, 
i.e.\ the geometrically thin, optically thick accretion disc introduced by Shakura \& Sunyaev (1973). 
According to this model, the emitted SED can be fitted with a multicolor black body spectrum, 
that depends only on the central black hole mass $M_{\rm BH}$ and on 
the accretion rate $\dot M$, directly traced by the overall disc luminosity 
through $L_{\rm d}=\eta\dot Mc^2$. 

Calderone et al.\ (2013) studied in depth the method and applied it to a 
sample of radio--loud Narrow Line Seyfert 1s, deriving some interesting 
features of the model. 
Specifically, we are interested in the relation between the peak luminosity 
and the overall luminosity of a multicolor black body spectrum. 
They found that $L_{\rm d}=2\nu_{\rm p}L_{\nu_{\rm p}}$, and therefore 
if the peak is clearly visible in the observed SED, an estimate of $L_{\rm d}$ 
can be easily derived. 
This is the case for the three objects we are considering. 
Deriving from the visible peak emission the overall disc luminosity leaves 
a single free parameter to the accretion disc fitting process, i.e.\ the black hole mass. 
At this condition of peak visibility, the average uncertainty on the 
black hole mass estimate is a factor 2--2.5. 
If the peak is not visible, instead, the average uncertainty reaches 
a factor 3.5--4, i.e.\ similar to the uncertainties obtained applying virial methods  
(Vestargaard \& Peterson 2006). 

\begin{figure*}
\vskip -0.6 cm 
\hskip -0.2 cm
\psfig{figure=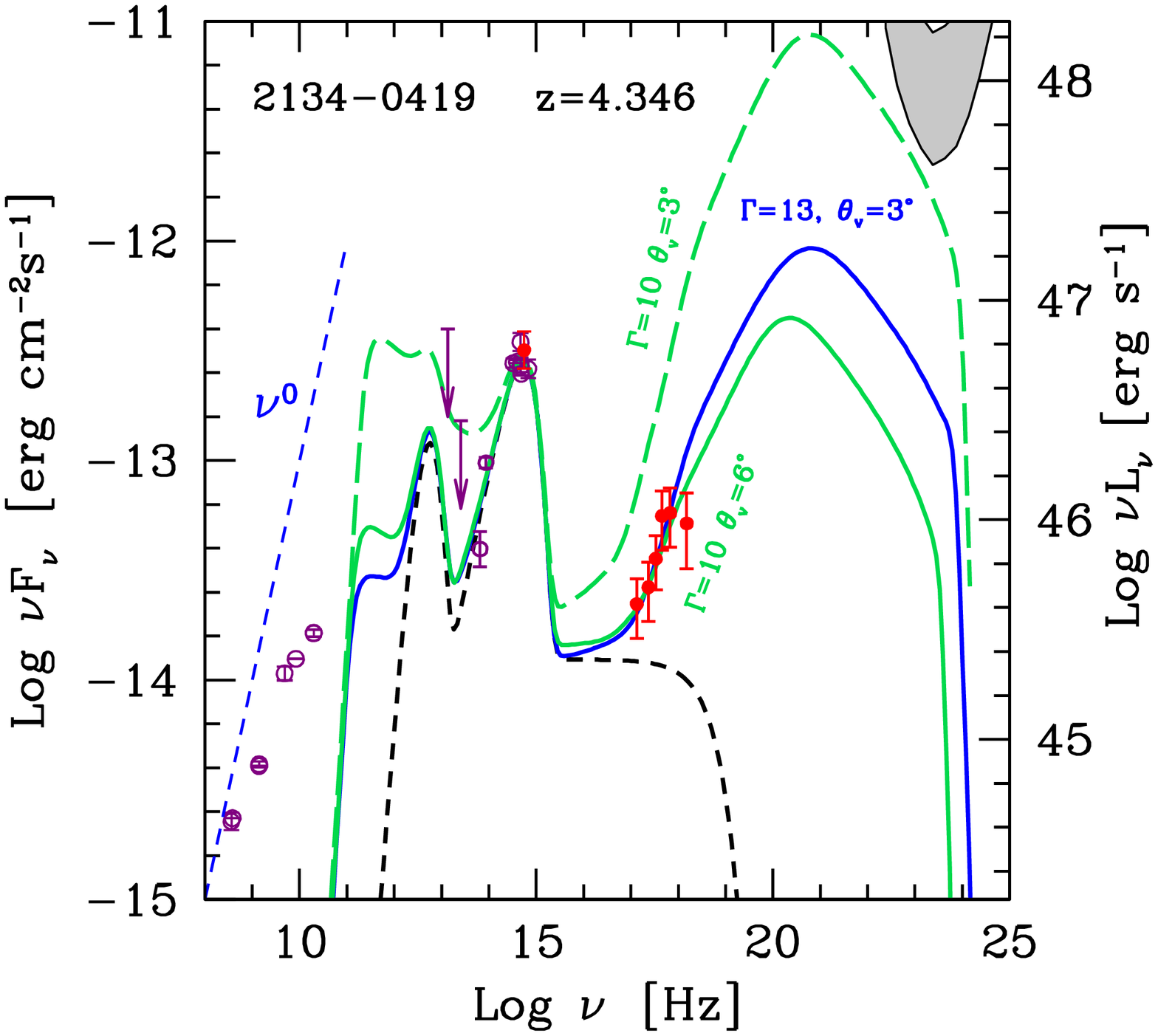,width=9cm,height=9cm}
\hskip -0.2 cm
\psfig{figure=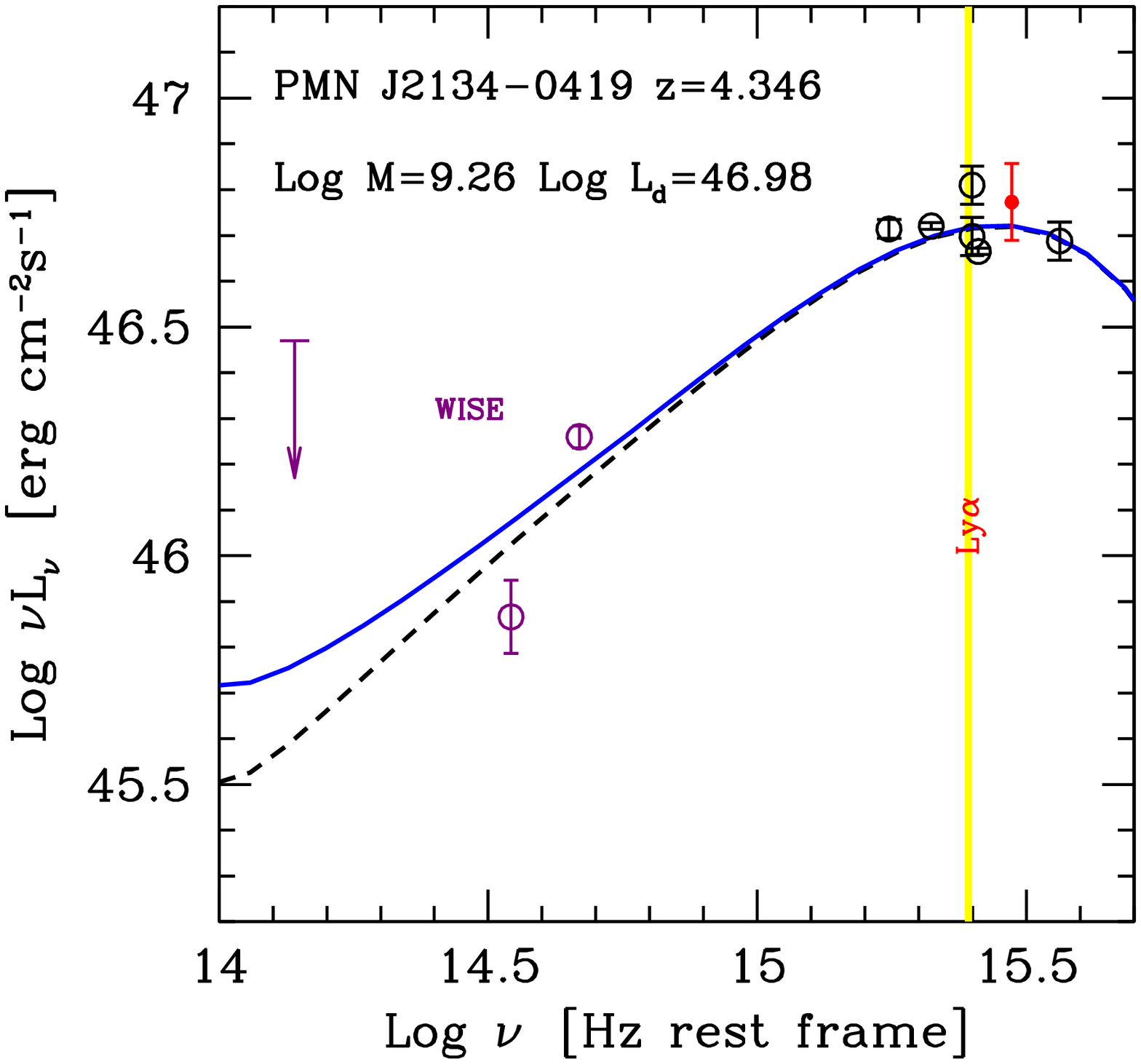,width=8.5cm,height=9cm}
\vskip -0.5 cm 
\caption{ {\it Left panel:} full SED of PMN J2134--0419. 
The blue solid line shows the best representation of the data. 
The green solid line is the fit made with largest possible viewing angle 
(second line of this source in Table \ref{para}), while the green dashed line 
is the model with the same parameters, but ``re--oriented" with a 
viewing angle $\theta_{\rm v}=3^\circ$. 
The black dashed line is the thermal emission from accretion disc, torus and corona. 
The blue dashed line is the extrapolated $F(\nu)\propto\nu^0$ from the observed 
radio data at lowest frequencies.  
Red data point are the new {\it Swift}/XRT and UVOT observations, 
purple and/or green data points are archival data. 
The curved grey stripe corresponds to the sensitivity of {\it Fermi}/LAT 
after 5 years of operations ($5\sigma$). 
{\it Right panel:} zoom of the SED on the IR--optical--UV 
wavelength range, to display the accretion disc fitting performed 
to measure the central black hole mass ($\log (M_{\rm BH}/M_\odot)=9.26$). 
The blue solid line shows the best representation of the data, purple data and 
upper limit are from WISE All--Sky Catalog, black empty 
data point are archive data, while the red point is the 
UVOT detection. 
}
\label{2134}
\end{figure*}

The optical--IR complete data coverage of SDSS J222032.50+002537.5 
and SDSS J142048.01+120545.9 is achieved thanks to SDSS spectra, 
the IR photometry from the Wide--field Infrared Explorer (WISE\footnote{
Data retrieved from \url{http://irsa.ipac.caltech.edu/}
}; Wright et al.\ 2010) 
and the optical--IR photometry obtained from the Gamma--Ray Burst 
Optical--Near Infrared Detector (GROND; Greiner et al.\ 2008). 
This allowed us to derive for both black hole mass estimates of 
$\log (M_{\rm BH}/M_\odot)=9.30$ in Sbarrato et al.\ (2013a). 
PMN J2134--0419, instead, was not included in that observational campaign, 
and thus lacks GROND data. 
Nevertheless, the peak of the accretion disc emission is visible also in this case 
(see Fig.\ \ref{2134}, right panel), 
and we could derive a black hole mass estimate for this object following the 
same method. 
We obtain a mass of $\log (M_{\rm BH}/M_\odot)=9.26$, confirming 
that this blazar candidate hosts an extremely massive black hole. 
The good spectral coverage of the peak of the disk emission
limits the uncertainties on the mass value to a factor 2.
Note that optical and IR data are not simultaneous, but we do not expect 
strong variations in this wavelength range 
(repeated GROND observations of the analogous object B2 1023+25 
did not show variability at all, Sbarrato et al.\ 2013b).

\begin{figure*}
\vskip -0.6 cm 
\hskip -0.2 cm
\psfig{figure=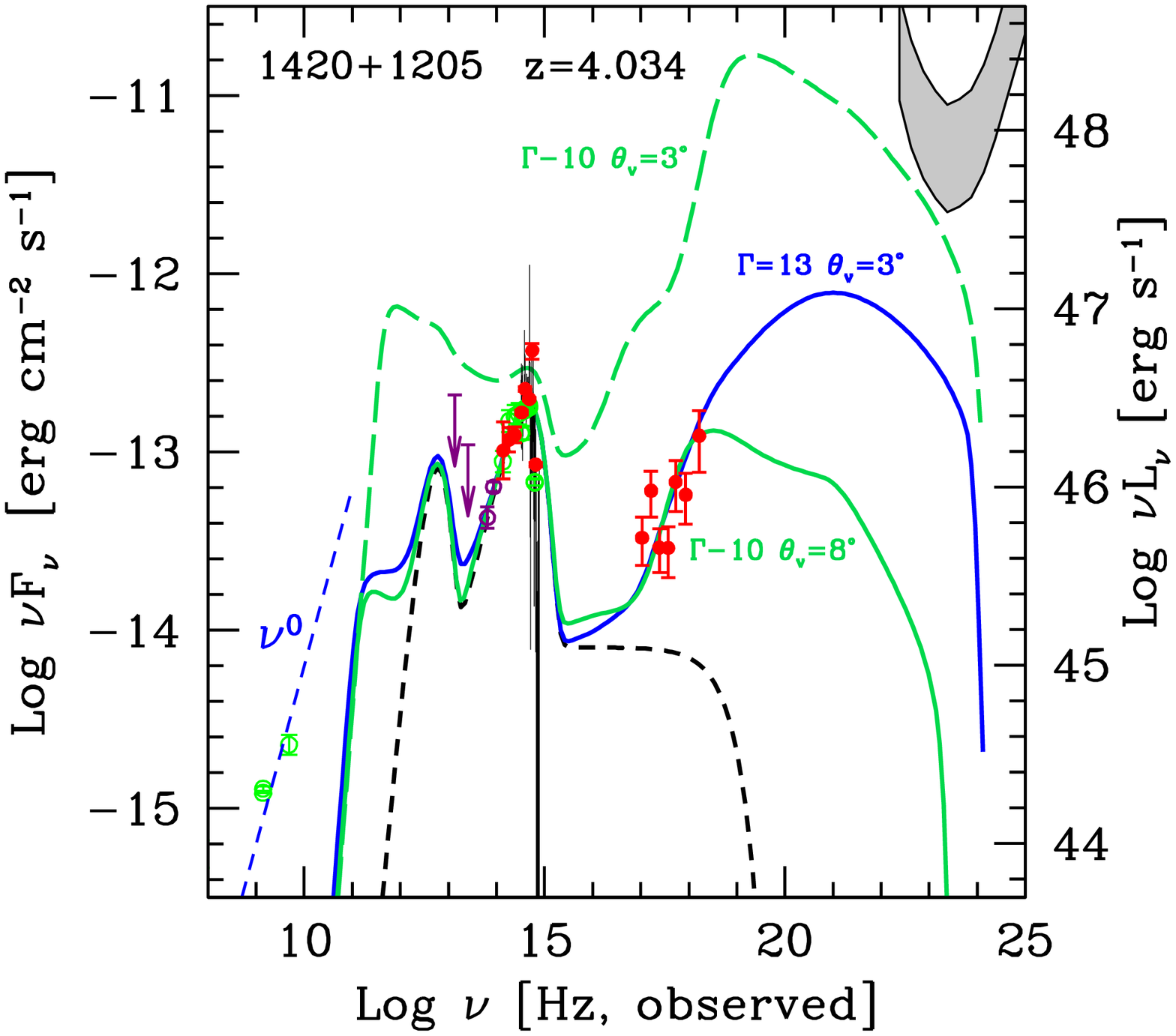,width=9cm,height=9cm} 
\hskip -0.2 cm
\psfig{figure=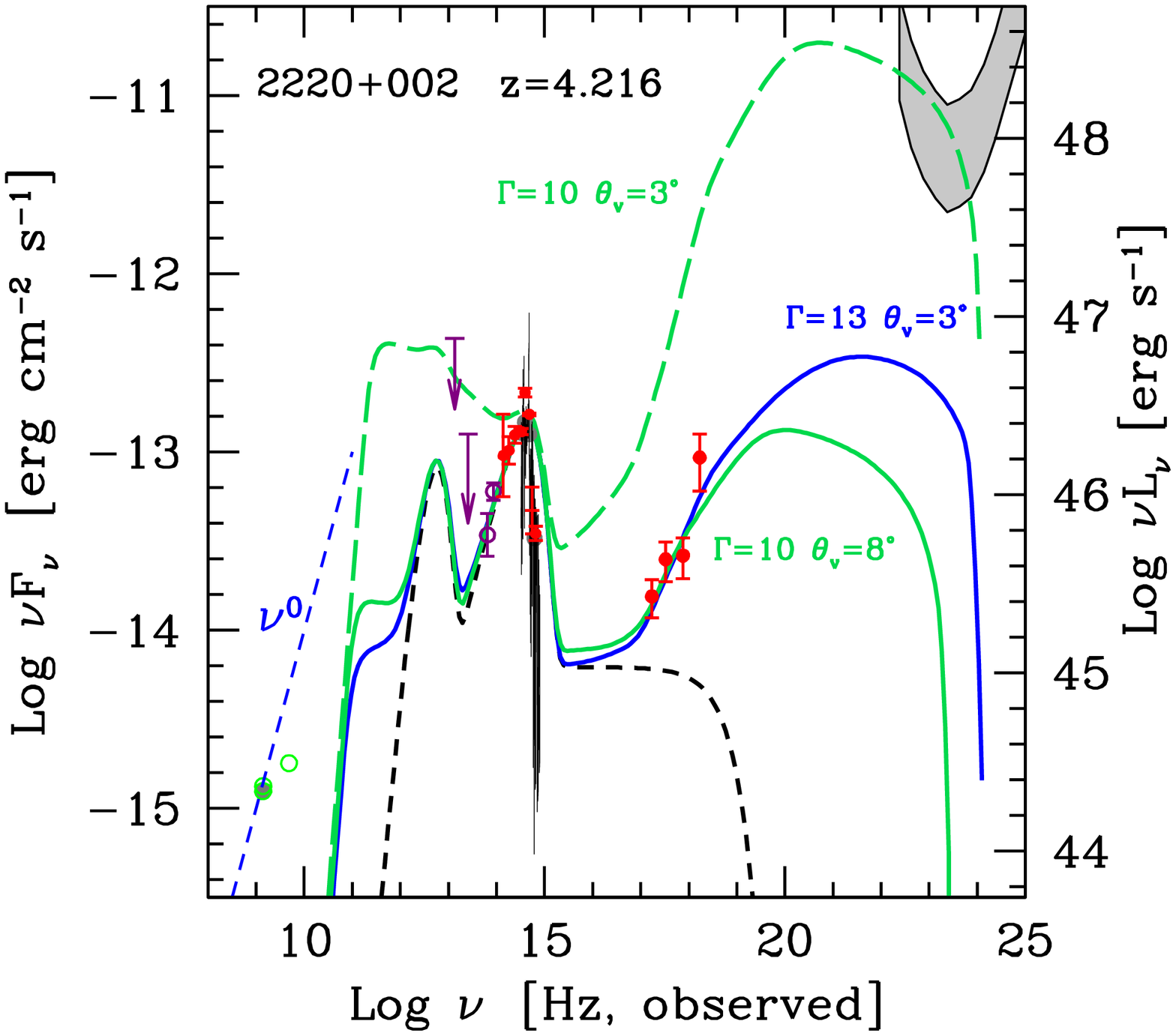,width=9cm,height=9cm}
\vskip -0.5 cm 
\caption{ SEDs of the two sources from the sample in Sbarrato et al.\ (2013a), 
observed by {\it Swift}/XRT: SDSS J142048.01+120545.9 ({\it left panel}) 
and SDSS J222032.50+002537.5 ({\it right panel}). 
In both figures, lines and data are as in Figure \ref{2134}, left panel. 
}
\label{1420}
\end{figure*}

\section{Observations and Data Analysis}
\label{data}

The analysis of the data from the X-Ray Telescope (XRT, Burrows et al. 2005) 
and UltraViolet Optical Telescope (UVOT, Roming et al. 2005) onboard the 
{\it Swift} satellite has been done by using HEASoft v 6.16 and the CALDB updated 
on 2014 September 4 and by following standard procedures as described 
e.g.\ in Sbarrato et al. (2012). 
Because of the low statistics, X--ray data were analysed by using unbinned likelihood (Cash 1979). 

Table \ref{xrt} shows the parameters of our analysis, with effective exposure, 
Galactic absorption, observed flux and photon index of the power law model 
for each source, along with the observation IDs.

\begin{table*} 
\centering
\begin{tabular}{llllllllllllll}
\hline
\hline
Name   &$z$ &$R_{\rm diss}$ &$M$ &$R_{\rm BLR}$ &$P^\prime_{\rm i}$ &$L_{\rm d}$ &$B$ &$\Gamma$ &$\theta_{\rm v}$
    &$\gamma_{\rm b}$ &$\gamma_{\rm max}$ &$s_1$  &$s_2$  \\
~[1]      &[2] &[3] &[4] &[5] &[6] &[7] &[8] &[9] &[10] &[11] &[12] &[13]  &[13] \\
\hline   
SDSS J142048.01+120545.9  &4.034 &360 (600) &2e9 &725 &6e--3 &53 (0.18) &2.6 &13 &3 &100 &3e3 &1   &2.5\\
~          &      &360 (600) &2e9 &725 &0.2   &53 (0.18) &3.4 &10 &8 &10  &3e3 &--1 &2.5\\
\hline 
PMN J2134--0419 &4.346 &432 (800) &1.8e9 &972 &7e--3 &95 (0.35) &2.9 &13 &3 &70  &4e3 &0   &2.6 \\
~          &      &540 (1e3) &1.8e9 &972 &0.08  &95 (0.35) &3.0 &10 &6 &70  &4e3 &--1 &2.6 \\
\hline 
SDSS J222032.50+002537.5   &4.216 &360 (600) &2e9   &671 &3e--3 &45 (0.15) &2.4 &13 &3 &100 &3e3 &1   &2.2 \\
~          &      &540 (900) &2e9   &671 &0.2   &45 (0.15) &2.1 &10 &8 &40  &3e3 &--1 &2.2 \\
\hline
\hline 
\end{tabular}
\vskip 0.4 true cm
\caption{List of parameters used to construct the theoretical SED.
Col. [1]: name;
Col. [2]: redshift;
Col. [3]: dissipation radius in units of $10^{15}$ cm and (in parenthesis) in units of \sc\ radii;
Col. [4]: black hole mass in solar masses;
Col. [5]: size of the BLR in units of $10^{15}$ cm;
Col. [6]: power injected in the blob calculated in the comoving frame, in units of $10^{45}$ erg s$^{-1}$; 
Col. [7]: accretion disk luminosity in units of $10^{45}$ erg s$^{-1}$ and
        (in parenthesis) in units of $L_{\rm Edd}$;
Col. [8]: magnetic field in Gauss;
Col. [9]: bulk Lorentz factor at $R_{\rm diss}$;
Col. [10]: viewing angle in degrees;
Col. [11] and [12]: break and maximum random Lorentz factors of the injected electrons;
Col. [13] and [14]: slopes of the injected electron distribution [$Q(\gamma)$] below and above $\gamma_{\rm b}$;
The total X--ray corona luminosity is assumed to be in the range 10--30 per cent of $L_{\rm d}$.
Its spectral shape is assumed to be always $\propto \nu^{-1} \exp(-h\nu/150~{\rm keV})$.
}
\label{para}
\end{table*}
%

\section{Jet and orientation}
\label{jet}

With the X--ray data, we complete a good coverage of the SEDs of all our candidates. 
Figures \ref{2134} and \ref{1420} show the SEDs of the 3 blazar candidates. 
The red points in the X--ray band are the new {\it Swift}/XRT data. 
All the other data points are described in the captions. 
The grey stripes show the sensitivity limit of {\it Fermi}/LAT. 
We fitted the observed data with the one--zone leptonic model described in 
Ghisellini \& Tavecchio (2009). 
In this model, relativistic electrons in the jet emit by synchrotron and IC processes, 
and their distribution is derived through a continuity equation in which we 
assume continuous injection, radiative cooling, possible pair production and emission. 
The particle distribution responsible for the
emission is calculated to occur at a time $R/c$ after the injection, 
where $R$ is the size of the emitting region, located at a distance 
$R_{\rm diss}$ from the central engine. 
We observe the jet under a a viewing angle $\theta_{\rm v}$ from the jet axis. 
Besides 
the jet emission, we take into account the emission from the 
accretion disc (particularly important to derive the black hole mass), 
the dusty torus and the hot thermal corona surrounding the disc itself. 

Fitting the overall SED, we obtain a set of parameters 
describing the source and its emitting condition. 
Two characterizing parameters are the viewing angle $\theta_{\rm v}$ 
and the bulk Lorentz factor $\Gamma$ of the emitting region. 
For each of our candidates, we present two models, summarized in 
Table \ref{para}. 
The best representation of the data is shown by the blue lines 
in Figures \ref{2134} and \ref{1420}. 
In all the three cases, this ``best fit'' describes a source seen under a viewing angle 
smaller than the jet beaming angle, i.e.\ $\theta_{\rm v}<1/\Gamma$. 
This allows to classify the three candidates as blazars, according to our criterion. 
The corresponding sets of parameters are consistent with what found for other powerful blazars. 
The other sets of parameters describe models with the largest possible jet viewing angle 
consistent with our data. 
To reproduce the same X--ray and radio fluxes, we have to associate 
smaller Lorentz factors compared to the ``best fit'', along with the parameters in the second 
line of each source of Table \ref{para}. 
The models described by these different sets of parameters are represented by green solid lines 
in Figure \ref{1420}. 

As a consistency check, for each source we test how an object with this last 
set of parameters would look like if seen at $\theta_{\rm v}<1/\Gamma$, i.e.\ 
under a typical viewing angle for blazars $\theta_{\rm v}=3^\circ$. 
These ``re--oriented'' models are shown in the figures with the dashed green lines. 
In the cases of SDSS J142048.01+120545.9 and SDSS J222032.50+002537.5 (Figure \ref{1420}), 
the resulting SEDs are extremely luminous in the X--rays.  
Although similar luminosities 
have already been observed in some known high--$z$ blazars, 
such as GB 1428+4217 ($z=4.72$; Worsley et al.\ 2004) 
and RX J1028--0844 (Zickgraf et al.\ 1997; Yuan et al.\ 2000), 
there would be some problems with the radio luminosity. 
If we extrapolate from the existing radio data of the two sources a slope of $F_\nu\propto\nu^0$ 
(i.e.\ the expected flat spectrum for a blazar), the ``re--oriented model'' would  
exceed that limit.
Moreover, the ``re--oriented'' version of SDSS J222032.50+002537.5 would 
also be detected by {\it Fermi}/LAT in the $\gamma$--rays, while so far no such 
object has been detected by LAT ({\it Fermi}/LAT sensitivity 
is shown in Fig. \ref{2134} and \ref{1420} with the grey shaded area).
These problems lead us to discard this solution, preferring the one with a small viewing angle
and a large $\Gamma$: 
if such objects existed, they would have been so luminous that they would already 
have been detected, and they were not. 
Therefore 
SDSS J142048.01+120545.9 and SDSS J222032.50+002537.5 can be classified as 
blazars, both with $\Gamma=13$ and $\theta_{\rm v}=3^\circ$. 

We repeat the fitting and the analysis on the third candidate, PMN J2134--0419. 
In this case, the largest angle solution gives $\theta_{\rm v}=6^\circ$. 
We repeated the consistency check studying a blazar with the same parameters as the large--angle 
version of PMN J2134--0419, but directed at smaller viewing angle ($\theta_{\rm v}=3^\circ$). 
In this case the re--oriented model is not far from the SED at $\theta_{\rm v}=6^\circ$, 
since the viewing angles are closer. 
The re--oriented model is also not completely unreasonable. 
For this object, therefore, we can state that the viewing angle is in the 
range $3^\circ-6^\circ$, with corresponding Lorentz factors of $13-10$.

To obtain a confirmation of the classification of these sources as blazars,
and possibly a more precise estimate of $\theta_{\rm V}$ and $\Gamma$, 
further observations would be useful:
the {\it NuSTAR} satellite 
could prove the behavior of these sources in the hard X--rays, clearly discriminating 
between a blazar solution ($\Gamma=13; \theta_{\rm v}=3^\circ$) and a slightly larger 
viewing angle associated with a smaller Lorentz factor. 
Furthermore,
the peak of the radio emission could be observed with the Atacama Large Millimiter/submillimiter Array (ALMA) 
or the Square Kilometre Array (SKA), 
that cover that wavelength range and has the sensitivity to detected the emission 
from this kind of sources.

\begin{figure}
\vskip -0.6 cm 
\hskip -0.3 cm
\psfig{figure=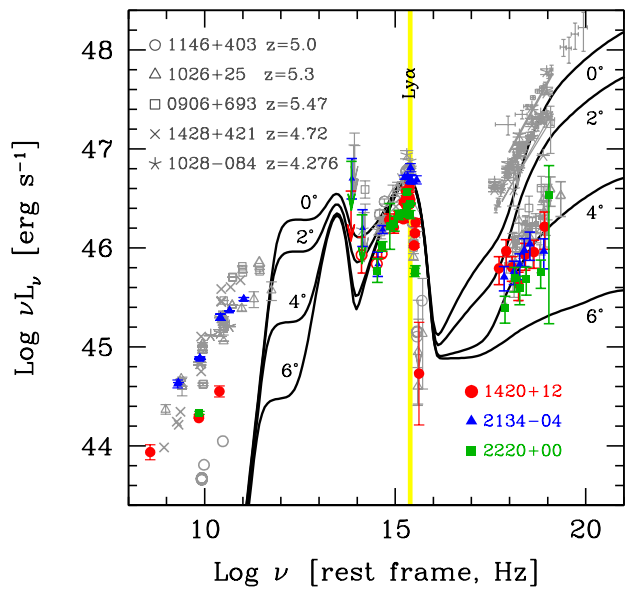,width=9cm,height=9cm}
\vskip -0.5 cm 
\caption{
Comparison of SDSS J142048.01+120545.9, SDSS J222032.50+002537.5 
and PMN J2134--0419 (red filled circles, green filled squares and blue 
filled triangles, respectively) with the three $z>5$ and the two 
most known $z>4$ blazars (data as labelled). 
GB 1428+4217 and RX J1028--0844 have X--ray spectra much more 
luminous than the other objects, as expected in case of objects found 
serendipitously. 
Superposed to the data (solid black lines),
SEDs of a typical powerful blazar oriented at 
different viewing angles (as labelled). 
Note how the observed flux from the external Compton emission is strongly 
dependent on the viewing angle (Dermer 1995), 
more than the corresponding synchrotron radiation.
This shows why the X--ray data help in finding the correct
viewing angle.
}
\label{z4l}
\end{figure}

\begin{figure*}
\vskip -0.6 cm 
\hskip -0.3 cm
\psfig{figure=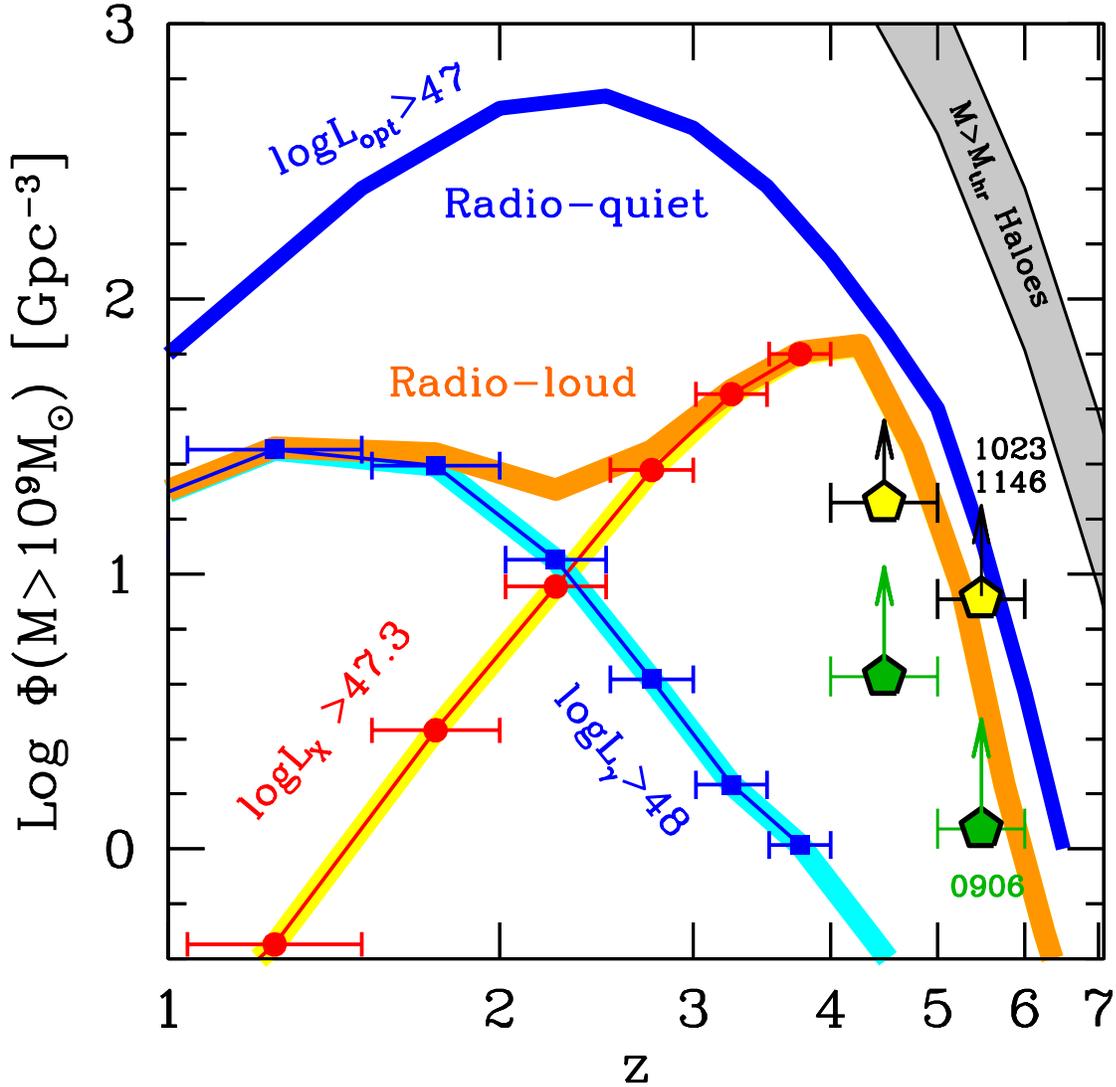,width=17cm,height=17cm}
\vskip -1.2 cm 
\caption{
Comoving number density of supermassive black holes with 
$M_{\rm BH}>10^9M_\odot$ hosted in radio--quiet (blue line, 
derived from the luminosity function by Hopkins et al.\ 2007) 
and radio--loud quasars (orange line). 
The radio--loud density is obtained from blazar number densities, 
by multiplying them by $450=2\Gamma^2$ ($\Gamma=15$). 
Blue data and the light blue line are derived from the $\gamma$--ray 
luminosity function obtained by {\it Fermi}/LAT (Ajello et al. 2012). 
Red data points and the yellow line are derived from the [15--55keV] 
luminosity function from Ajello et al.\ (2009), modified as in Ghisellini et al.\ (2010). 
All the number density functions are derived by integrating the corresponding 
luminosity functions at luminosities larger than what labelled in Figure, to 
ensure that correspond to $M_{\rm BH}>10^9M_\odot$. 
Such a cut in luminosity selects objects that are the most luminous in 
their corresponding bands, other than the most massive. 
Green pentagons represent the state of the art before the beginning 
of our project, with 4 serendipitous blazars in the $4<z<5$ bin and the
single detection of Q0906+6930 at $z>5$. 
The yellow pentagons are instead the number densities derived from our results. 
In the redshift frame $5<z<6$ the data point is given by the two blazars 
we classified at $z>5$ (B2 1023+25 and SDSS J1146+403, both in the SDSS+FIRST region of the sky). 
At $4<z<5$ the new (yellow) lower limit is provided by the two already known high--$z$ blazars 
in the SDSS+FIRST survey (SDSS J083946.22+511202.8 and SDSS J151002.92+570243.3), 
along with the three classifications we perform in this work. 
Our results confirm the existence 
of an early peak ($z\sim4$) of black hole formation 
in jetted AGN, in contrast to the main formation epoch of massive radio--quiet 
quasars ($z\sim2.5$).
}
\label{fi}
\end{figure*}

\subsection{Comparison with known blazars} 
\label{comp}

To put these three objects in the wider picture of known high--redshift blazars, 
we compare them with the most studied objects. 
Fig. \ref{z4l} shows our three blazar candidates compared 
with the three blazars known at $z>5$ (Q0906+6930, B2 1023+25, SDSS J1146+403) 
and the two best studied blazars at $4<z<5$ (GB 1428+4217 and RX J1028--0844). 
It can be immediately noticed that GB 1428+4217 and RX J1028--0844 
are much more luminous in the X--rays than the other blazars considered. 
This is likely connected to the fact that those two objects where discovered serendipitously, 
and therefore they must be the most luminous sources of their kind. 
Our systematic approach, instead, chases less extreme blazars, since 
we start our selection from their optical features, that trace the disc accretion 
and not the jet, and we selected them from a complete optical quasar catalog. 

To understand how our fits are sensitive to the viewing angle, Figure \ref{z4l} also 
shows how a SED with fixed parameters can vary with small variations 
in the viewing angle. 
The difference between the extreme blazars GB 1428+4217 and RX J1028--0844 
and our three blazar candidates (along with the three $z>5$ confirmed blazars) 
could be just due to a difference in orientation of $4^\circ$. 
Strong beaming is responsible for such extreme variations. 
The beaming pattern produced by a relativistically moving source that emits for 
external Compton (as expected in FSRQs as our high--$z$ blazars) is in fact 
strongly dependent on the viewing angle, through the beaming factor 
$\delta=[\Gamma(1-\beta\cos\theta_{\rm v})]^{-1}$. 
Dermer (1995) calculated that the observed flux of such a relativistic emitting region goes as
$\delta^{4+2\alpha}$, where $\alpha$ is the energy spectral index of the emitted radiation. 
This dependence is stronger than the one 
characterizing synchrotron emission, which is $\propto\delta^{3+\alpha}$. 
This determines the variations of radio emission as a function of viewing angle\footnote{
In our
one zone model the radio emission is self--absorbed, but the expected 
change of the flux as we change $\theta_{\rm v}$
can be seen through the far IR synchrotron emission that shows 
a smaller range of variation than the hard X--rays.
}. 
It is worth to notice that this is the reason why observations of high--energy humps are  
necessary to classify a source as a blazar, while the radio luminosity alone is not enough. 
The high--energy emission is in fact more sensitive to viewing angle variations than the synchrotron.

\section{Two epochs for black hole formation?}
\label{fi_sec}

Our three blazars join the other two blazars already known in the SDSS+FIRST region of the sky in
the redshift bin $4<z<5$ and with a black hole mass $M_{\rm BH}>10^9 M_\odot$:
SDSS J083946.22+511202.8 and SDSS J151002.92+570243.3 (Sbarrato et al.\ 2013a).
They allow us to infer the existence of a large number of jetted quasars analogous to the 5 blazars. 
Since they are all observed with $\theta_{\rm v}<1/\Gamma$, each of them 
traces the presence of $\sim2\Gamma^2=338(\Gamma/13)^2$ quasars with $M_{\rm BH}>10^9M_\odot$. 
Since the SDSS+FIRST survey covers 8770 deg$^2$, the 5 blazars imply
that over the whole sky there must exist $\sim7700$ jetted AGN with similar intrinsic properties,
namely similar black hole masses.
The comoving volume in the redshift frame $4<z<5$ is $\sim425$ Gpc$^3$, therefore 
we can conclude that there must be at least 18 radio--loud AGN per Gpc$^3$ with  
masses $M_{\rm BH}>10^9M_\odot$, hosted in jetted systems.

How does this conclusion fit in the current paradigm of supermassive black holes in the early Universe? 
Fig. \ref{fi} shows the comoving number density of extremely massive black holes 
($M_{\rm BH}>10^9M_\odot$) hosted by radio--quiet (blue line, derived as in Ghisellini et al.\ 2010 
from the mass function in Hopkins et al.\ 2007) and radio--loud AGN [orange line, derived from 
{\it Fermi}/LAT (Ajello et al. 2012) and {\it Swift}/BAT (Ajello et al. 2009) blazar luminosity 
functions as in Ghisellini et al.\ 2010].  
Note that at $z>4$ the comoving number density of jetted quasars is no more supported by 
data from the two blazar surveys ({\it Fermi}/LAT and {\it Swift}/BAT). 
Before the beginning of our systematic search of high--redshift blazar candidates 
(see Ghisellini et al.\ 2010; 2013), 
the serendipitous blazars known at $z>4$ (green pentagons) could not provide 
sufficient statistics to continue the calculation of comoving number density. 
For $z>4$ the density was assumed to decrease exponentially,  
as the corresponding one for radio--quiet objects. 

Nevertheless, a hint of different density distributions between jetted and non--jetted objects 
was already visible in Ghisellini et al.  (2010) and Ghisellini et al. (2013). 
The two yellow pentagons 
in Fig. \ref{fi} are the (all--sky) number densities derived from the 
5 blazars at $4<z<5$ contained in the SDSS+FIRST sky area
(3 from this work and 2 from Sbarrato et al.\ 2013a) 
and the two blazars we classified at $z>5$ 
(B2 1023+25 at $z=5.3$, Sbarrato et al.\ 2012, 2013b; SDSS~J1146+403 at $z=5$, 
Ghisellini et al.\ 2014). 
Our observations clearly push towards an interesting conclusion: the density of extremely massive 
black holes hosted in jetted systems peak at least around $z\sim4$, while the non--jetted 
systems peak at $z\sim2-2.5$. 
This suggests two different epochs of SMBH formation, and 
the black holes that grow developing a jet seem to be born earlier, and/or to grow faster. 

The presence of a jet in AGN is commonly linked to high values of black hole spin. 
This does not facilitate a fast accretion, according to the common knowledge. 
Maximally spinning black holes (i.e.\ with dimensionless spin values $a\sim0.998$) 
accrete from accretion discs that are thought to be more efficient 
radiators ($\eta=0.3$; Thorne 1974). 
Spending energy in radiation makes the accretion of matter on the black hole 
much less efficient, slowing down the accretion process. 
As explained in Ghisellini et al.\ (2013), in fact, a spinning black hole accreting at 
Eddington rate would need 3.1 Gyr to grow from a seed of $100M_\odot$ to $10^9M_\odot$
(ignoring black hole merging).
This would imply that such massive black holes should not be visible at $z>2.1$, 
while their preferential formation epoch seems to be around $z\sim4$. 
In Ghisellini et al.\ (2013) some options for a faster accretion in presence 
of a jet are explored. 
The available energy, in fact, is not all radiated away, but contributes to 
amplifying the magnetic field and thus launching the jet. 
Considering this, the accretion is faster, but black holes with $M_{\rm BH}>10^9M_\odot$ 
are still hard to form before $z\sim4-5$.

\section{Conclusions}
\label{concl}

In this work, we observed with {\it Swift}/XRT 
three blazar candidates contained in the SDSS+FIRST survey,
having redshifts between 4 and 5 and black hole masses
exceeding $10^9 M_\odot$.
We can classify SDSS J142048.01+120545.9, SDSS J222032.50+002537.5 and PMN J2134--0419
as blazars, thanks to the their bright and hard X--ray spectrum.
The full SED fitting in fact requires bulk Lorentz factors
$\Gamma\sim 13$, and viewing angles $\theta_{\rm v}\sim 3^\circ$. 

These three newly classified blazars join 
the other two already known in the same region of the sky,
same redshift bin, and black hole mass exceeding $10^9 M_\odot$.
We can then infer
the presence of at least $\sim$18 extremely massive black holes 
per Gpc$^3$ hosted in jetted systems in the redshift frame $4<z<5$, i.e.\ $\sim7700$ 
in the whole sky. 

Populating the high--redshift density function of jetted AGN provides interesting 
constraints to supermassive black hole formation and evolution models. 
Our results confirm the existence
of two different formation epochs for supermassive black holes: 
the most extremely massive objects seem to form earlier, at $z\sim4$, in jetted 
than in non--jetted systems, whose peak is at $z\sim2.5$.

\section*{Acknowledgments}
We thank the anonymous referee for useful comments. 
This research made use of the NASA/IPAC Extragalactic Database (NED) 
and of the data products from the Wide--field Infrared Survey Explorer, 
which are operated by the Jet Propulsion Laboratory, Caltech, funded by 
the National Aeronautics and Space Administration. 
Part of this work is based on archival data, software and online services 
provided by the ASI Science Data Center (ASDC). 
We made also use of data supplied by the UK {\it Swift} Science Data Centre 
at the University of Leicester.


\end{document}